\documentclass[fleqn,12pt]{article}
\usepackage{latexsym,amsmath}      
                                                  
\newcommand{\p}{\partial}

\newcommand{\ds}{\displaystyle}
\newtheorem{theo}{Theorem}
\newcommand{\bt}{\begin{theo}}
\newcommand{\et}{\end{theo}}
\newtheorem{rem}{Note}
\newcommand{\br}{\begin{rem}}
\newcommand{\er}{\end{rem}}
\newtheorem{corol}{Corollary}
\newcommand{\bcor}{\begin{corol}}
\newcommand{\ecor}{\end{corol}}

\begin{document}

\begin{flushleft}
\Large \bf 
Continuity Equation in Nonlinear Quantum\\
 Mechanics  and the Galilei Relativity Principle
\end{flushleft}

\bigskip

\noindent
{\bf Wilhelm I. FUSHCHYCH~$^{\dag}$
and Vyacheslav M. BOYKO~$^{\ddag}$}

\bigskip

\noindent
{\it Institute of Mathematics of the National Ukrainian Academy of Sciences,\\
3 Tereshchenkivska Str. 3, Kyiv 4, 01601 Ukraina}

\medskip

\noindent
$^{\dag}$~URL: {\tt http://www.imath.kiev.ua/\~{}appmath/wif.html}

\noindent
$^{\ddag}$~URL: {\tt http://www.imath.kiev.ua/\~{}boyko/} \ \ 
E-mail: {\tt boyko@imath.kiev.ua}

\bigskip

\begin{abstract}
\noindent
Classes of the nonlinear  Schr\"odinger-type  equations
compatible with the Galilei relativity principle are described.
Solutions of these equations satisfy   the continuity equation.
\end{abstract}

\noindent
The continuity equation is one of the most fundamental equations of quantum
mechanics
\begin{equation} \label{1}
{\p \rho \ \over \p t} +\vec \nabla \cdot \vec j=0.
\end{equation}
Depending on definition of $\rho$ (density) and $\vec j= (j^1,\ldots,
j^n)$ (current), we can construct essentially different quantum
mechanics with different  equations of motion, which are distinct
from classical linear Schr\"odinger, Klein--Gordon--Fock, and Dirac
equations.

In this paper we describe  wide classes of the nonlinear
Schr\"odinger-type equations compatible with
the Galilei relativity principle and their solutions
 satisfy the continuity equation.

\noindent
{\bf 1.} At the beginning we study a symmetry of the continuity equation
considering  $(\rho,  \vec j)$ as dependent variables
related by (\ref{1}).

\vspace*{-2mm}

\bt
The invariance algebra of equation (\ref{1}) is an infinite-dimensional
algebra with basis operators
\begin{equation} \label{2}
X= \xi^\mu(x) {\p \ \over \p x_\mu} + \Bigl(a^{\mu\nu}(x) j^\nu+ b^\mu(x)
\Bigr){\p \ \over \p j^\mu}\; ,
\end{equation}
where $j^0\equiv \rho$; $\xi^\mu(x)$ are arbitrary smooth functions;
$x =(x_0=t, x_1, x_2, \ldots, x_n)\in {\bf R}^{n+1}$;
$\ds a^{\mu\nu}(x) ={\p \xi^\mu \over \p x_\nu}- \delta_{\mu\nu}\left({\p
\xi^i \over \p x_i}+C\right)$; $C=\,{\rm const}$,
$\delta_{\mu\nu}$ is the Kronecker delta; $\mu, \nu, i = 0,1, \ldots, n$,
$\Bigl(b^0(x), b^1(x), \ldots, b^n(x)\Bigr)$ is an arbitrary solution of
equation~(\ref{1}).
\et
Here and below we imply summation over repeated indices.

\vspace*{-2mm}

\bcor
The  generalized Galilei algebra \cite{1}
\begin{equation} \label{3}
AG_2(1,n)= < P_\mu, J_{ab}, G_a, D^{(1)}, A>
\end{equation}
is a subalgebra of  algebra (\ref{2}).
\ecor

\vspace*{-2mm}

\bcor
The conformal   algebra \cite{1}
\begin{equation} \label{4}
AP_2(1,n)= AC(1,n)=< P_\mu, J_{ab}, J_{0a}, D^{(2)}, K_\mu>
\end{equation}
is a subalgebra of  algebra (\ref{2}).
\ecor

We  use the following designations in (\ref{3}) and (\ref{4})
\[
\arraycolsep=0pt\begin{array}{l}
P_\mu =\p_\mu, \quad J_{ab}= x_a\p_b- x_b\p_a +
j^a\p_{j^b}- j^b\p_{j^a}, \quad (a<b)
\vspace{2mm}\\
G_a= x_0\p_a +\rho\p_{j^a}, \quad
J_{0a} = x_a\p_0+x_0\p_a +j^a\p_\rho +\rho\p_{j^a},
\vspace{2mm}\\
D^{(1)} =2x_0\p_0 +x_a\p_a -n\rho\p_\rho-(n+1)j^a\p_{j^a}, \quad
D^{(2)} =x_\mu\p_\mu  -n\rho\p_\rho-nj^a\p_{j^a},
\vspace{2mm}\\
A= x^2_0\p_0+ x_0x_a\p_a -nx_0\rho\p_\rho+ (x_a\rho- (n+1)x_0j^a)\p_{j^a},
\vspace{2mm}\\
K_\mu= 2x_\mu D^{(2)} -x_\nu x^\nu g_{\mu i}\p_i
- 2x^\nu S_{\mu\nu}, \quad
 S_{\mu\nu}= g_{\mu i}j^\nu \p_{j^i}-g_{\nu i} j^\mu \p_{j^i},
\vspace{2mm}\\
 g_{\mu\nu}=\left\{
\arraycolsep=0pt\begin{array}{rl}
1, & \ \ \mu =\nu =0 \\
-1, & \ \ \mu =\nu \not=0\\
0, & \ \ \mu\not=\nu,
\end{array} \right.
\quad
\mu, \nu, i =0,1\ldots,n; \ a,b =1,2,\ldots, n.
\end{array}
\]
\bcor
The continuity equation satisfies the Galilei relativity principle
as well as the Lorentz--Poincare--Einstein relativity principle.
\ecor
Thus, depending on the definition of $\rho$ and $\vec j$, we come to
different quantum mechanics.
\\[1.5mm]
{\bf 2.} Let us consider the scalar complex--valued wave functions
and define $\rho$ and $\vec j$ in the following way
\begin{equation} \label{5}
\arraycolsep=0pt\begin{array}{l}
\rho = f(uu^*),
%\vspace{2mm}\\
\ \ \ds j^k = -\frac 12 i g(uu^*)\left( {\p u \over \p x_k} u^* -u {\p u^*
\over \p x_k} \right)+ {\p \varphi(uu^*) \over \p x_k}\; ,
\ \
k=1,2,\ldots, n.
\end{array}
\end{equation}
where $f, \ g, \ \varphi$ are arbitrary smooth functions,
$f\not= \,{\rm const}$, $g \not= 0$. Without loss of generality, we assume
that $f \equiv uu^*$.

Let us describe all functions $g(uu^*)$, $\varphi (uu^*)$ for
continuity equation (\ref{1}), (\ref{5}) to be compatible with the Galilei
relativity principle, defined by the following transformations:
\[
t \to t'= t, \qquad
x_a\to x_a'= x_a +v_a t.
\]
Here we do not fix transformation rules for the wave function $u$.

\bt
If $\rho$ and $\vec j$ are defined according to  formula (\ref{5}), then the
continuity equation (\ref{1}) is Galilei--invariant iff
\begin{equation} \label{6}
\arraycolsep=0pt\begin{array}{l}
\rho = uu^*,
%\vspace{2mm}\\
\ \ \ds j^k = -\frac 12 i \left( {\p u \over \p x_k} u^* -u {\p u^*
\over \p x_k} \right)+ {\p \varphi(uu^*) \over \p x_k} \; ,
\ \
k=1,2,\ldots, n.
\end{array}
\end{equation}
The corresponding generators of  Galilei transformations  have the form
\[
G_a= x_0\p_a + ix_a\left (u\p_u- u^*\p_{u^*}\right), \quad a=1,2,\ldots, n.
\]
\et
If in (\ref{6})
\begin{equation} \label{7}
\varphi = \lambda \; uu^*, \qquad \lambda =\,{\rm const},
\end{equation}
then the continuity equation (\ref{1}), (\ref{6}), (\ref{7}) coincides
with the Fokker--Planck equation
\begin{equation} \label{71}
{\p \rho \ \over \p t} +\vec \nabla \cdot \vec j + \lambda \Delta \rho=0,
\end{equation}
where
\begin{equation} \label{72}
\arraycolsep=0pt\begin{array}{l}
\rho = uu^*,
%\vspace{2mm}\\
\quad
\ds j^k = -\frac 12 i \left( {\p u \over \p x_k} u^* -u {\p u^* \over \p
x_k} \right)\; , \quad k=1,2,\ldots, n.
\end{array}
\end{equation}
The continuity equation  (\ref{1}), (\ref{6}), (\ref{7})
 was considered in  \cite{2,6}.

Let us invistigate the symmetry of the nonlinear Schr\"odinger equation
\begin{equation} \label{8}
i u_0+ \frac 12 \Delta u+i {\Delta \varphi (uu^*) \over 2uu^*} u=
F\left( uu^*, (\vec \nabla (uu^*))^2, \Delta(uu^*)\right)u,
\end{equation}
where $F$ is an arbitrary real smooth function.

For the solutions of equation (\ref{8}),  equation (\ref{1}), (\ref{6})
is satisfied and is compatible with the Galilei
relativity principle.
Schr\"odinger  equations in the form of (\ref{8}), when $\varphi(uu^*)=
\lambda uu^*$ for fixed  function  $F$, were considered in \cite{1}--\cite{10}.

In  terms of the phase and amplitude $\Bigl( u =R\exp(i \Theta)\Bigr)$,
equation (\ref{8})  has the form
\begin{equation} \label{9}
\arraycolsep=0pt\begin{array}{l}
\ds R_0 +R_k\Theta_k +\frac 12 R\Delta \Theta +{1 \over 2R} \Delta \varphi
=0,
\vspace{2mm}\\
%\quad
\ds \Theta_0 +\frac 12 \Theta^2_k -{1\over 2R}\Delta R +
F\left(R^2, \left(\vec \nabla \left(R^2\right)\right)^2, \Delta R^2\right)= 0.
\end{array}
\end{equation}
\bt
The maximal invariance algebras for  system (\ref{9}), if $F=0$, are
the following:
\begin{equation} \label{10}
1. \quad <P_\mu, J_{ab}, Q, G_a, D>
\end{equation}
when $\varphi$ is an arbitrary function;
\begin{equation} \label{11}
2. \quad <P_\mu, J_{ab}, Q, G_a, D, I, A>
\end{equation}
when $\varphi =\lambda R^2$, $ \lambda =\,{\rm const}$.
\et
In (\ref{10}) and (\ref{11}) we use the following designations:
\begin{equation} \label{111}
\arraycolsep=0pt\begin{array}{l}
 P_\mu = \p_\mu, \qquad J_{ab} =x_a\p_{x_b} -x_b\p_{x_a}, \qquad a< b,
\vspace{2mm}\\
G_a= x_0\p_{x_a} +i x_a\p_\Theta, \quad Q= \p_\Theta, \quad D =
2x_0\p_{x_0} +x_a\p_{x_a}, \quad I =R\p_R, \vspace{2mm}\\
\ds A =x^2_0\p_{x_0} +x_0x_a\p_{x_a} -\frac n2 x_0R\p_R +
\frac 12 x^2_a\p_\Theta, \vspace{2mm}\\
\mu =0,1, \ldots, n; \qquad  a,b=1,2, \ldots,n.
\end{array}
\end{equation}
Algebra (\ref{11}) coincides with the invariance algebra of the linear
Schr\"odinger equation.

\bcor
System (\ref{9}), (\ref{7}) is invariant with respect to algebra
(\ref{11}) if
\[
F= R^{-1} \Delta R \; N\left({R \Delta R \over (\vec \nabla R)^2}\right),
\]
 where $N$ is an arbitrary real smooth function.
\ecor
{\bf 3.} Let us consider a more general system than (\ref{8})
\begin{equation} \label{12}
i u_0+ \frac 12 \Delta u= (F_1+ i F_2)u,
\end{equation}
where $F_1, F_2$ are arbitrary real smooth functions,
\begin{equation} \label{12'}
F_m= F_m\left( uu^*, (\vec \nabla (uu^*))^2, \Delta(uu^*)\right)u, \qquad
m=1,2.
\end{equation}

The structure of functions $F_1, F_2$ may be
described in  form (\ref{12'}) by virtue of conditions for
system (\ref{12}) to be Galilei-invariant.

In  terms of the phase and  amplitude, equation (\ref{12}) has the
form
\begin{equation} \label{13}
\arraycolsep=0pt\begin{array}{l}
\ds R_0 +R_k\Theta_k +\frac 12 R\Delta \Theta - RF_2=0,
%\vspace{2mm}\\
\quad
\ds \Theta_0 +\frac 12 \Theta^2_k -{1\over 2R}\Delta R +F_1=0,
\end{array}
\end{equation}
where $F_m= F_m\left(R^2, \left(\vec \nabla \left(R^2\right)\right)^2,
\Delta R^2\right), \quad  m=1,2$.

\bt System (\ref{13}) is invariant with respect to the generalized Galilei
algebra \\
$AG_2(1,n) =<P_\mu, J_{ab}, G_a, Q, \widetilde D, A>$ if it has the form
\[
\arraycolsep=0pt\begin{array}{l}
\ds R_0 +R_k\Theta_k +\frac 12 R\Delta \Theta -
R^{1+4/n} \; M\left(
{(\vec \nabla R)^2\over R^{2+4/n}} \; ; {\Delta R \over R^{1+4/n}}\right)=0, \vspace{2mm}\\
\ds \Theta_0 +\frac 12 \Theta^2_k -{1\over 2R}\Delta R +
R^{4/n} \; N\left(
{(\vec \nabla R)^2\over R^{2+4/n}} \; ; {\Delta R \over R^{1+4/n}}\right)=0,
\end{array}
\]
where $N, M$ are arbitrary real smooth functions.
The basis operators of the algebra
\linebreak
$AG_2(1,n)$ are defined by (\ref{111}) and
$\ds \widetilde D = D-\frac n2 I$.
\et
\bt
System (\ref{13}) is invariant with respect to  algebra (\ref{11})
if it has the form
\begin{equation} \label{14}
\arraycolsep=0pt\begin{array}{l}
\ds R_0 +R_k\Theta_k +\frac 12 R\Delta \Theta -\Delta R\; M\left(
{R \Delta R \over (\vec \nabla R)^2}\right)=0, \vspace{2mm}\\
\ds \Theta_0 +\frac 12 \Theta^2_k -{1\over 2R}\Delta R +
{\Delta R \over R} N\left({R \Delta R \over (\vec \nabla R)^2}\right)=0,
\end{array}
\end{equation}
where $N, M$ are arbitrary real smooth functions.
\et

System (\ref{14}) written in  terms of the wave function has the form
\begin{equation} \label{15}
i u_0+ \frac 12 \Delta u=
{\Delta |u| \over |u| }
\left(N\left( {|u|\Delta |u| \over (\vec \nabla |u|)^2}\right)
+iM\left( {|u|\Delta |u| \over (\vec \nabla |u|)^2}\right)\right)u.
\end{equation}
Equation (\ref{15}) is equivalent to the following equation
\[
i u_0+ \frac 12 \Delta u=
{\Delta (uu^*) \over (uu^*) }
\left(\tilde N \left( {(uu^*)\Delta (uu^*) \over (\vec \nabla
(uu^*))^2}\right)
+i\tilde M\left( {(uu^*)\Delta (uu^*) \over (\vec \nabla
(uu^*))^2}\right)\right)u.
\]

Thus,  equation (\ref{14}) admits an invariance algebra which coincides
with the invariance algebra of the linear Schr\"odinger equation with the
arbitrary functions $M, N$.
\br
With certain particular $M$ and $N$ the symmetry of  system (\ref{14})
can be essentially extended. E.g.,  if in (\ref{14}) $\ds N =\frac 12$,
then the second equation of the system (equation for the phase) will
be the Hamilton--Jacobi equation \cite{5}.
\er
Let us consider some forms of the continuity equation (\ref{1}) for
equation (\ref{14}).

\noindent
\underline{Case 1.} If $M=0$, then for solutions of equation  (\ref{14})
 equation (\ref{1}) holds true,
where the density and  current can be  defined in the classical way
(\ref{72}).

\noindent
\underline{Case 2.} If $\ds \Delta R\; M = -\lambda
\Bigl (\Delta R + {
(\vec \nabla R)^2 \over R}\Bigr )$,  then for
solutions of equation (\ref{14}),  the continuity equation (\ref{1}),
(\ref{6}), (\ref{7}) (or the Fokker--Planck equation (\ref{71}),
(\ref{72})) is valid.

\noindent
\underline{Case 3.} If $M$ is arbitrary then for solutions of equation
(\ref{14}), the continuity equation is valid, where the density and
current can be  defined by the conditions
\[
\arraycolsep=0pt\begin{array}{l}
\rho = uu^*,
%\vspace{2mm}\\
\ \ \ \ds \vec \nabla \cdot \vec j = {\p \ \over \p x_k }\left(-\frac 12 i
\left( {\p u \over \p x_k} u^* -u {\p u^* \over \p x_k} \right) \right)
-2 |u| \Delta |u| \; M\left( {|u|\Delta |u| \over (\vec \nabla
|u|)^2}\right) \; .
\end{array}
\]

Thus, we constructed wide classes of the nonlinear  Schr\"odinger-type
equations which is invariant with respect to  algebra (\ref{11})
(maximal invariance algebra of the linear Schr\"odinger equation) and for
whose solutions  the continuity equation (\ref{1}) is valid.

\end{document}